\documentclass[11pt]{article}
\usepackage[dvipdfmx]{color}
\usepackage{graphicx}
\usepackage{xcolor}
  \usepackage{amsthm,amsfonts}
  \usepackage{amsmath}
\usepackage{mathabx}
\usepackage{slashed}
\usepackage{ulem}
\usepackage{hyperref}
\usepackage{cleveref}
\usepackage{cite}

\newcommand{\bea}   {\begin{eqnarray}}
\newcommand{\eea}   {\end{eqnarray}}
\def\zzg{${\mathbb Z}_2\times{\mathbb Z}_2$-graded }

\begin{document}
\renewcommand{\thefootnote}{\fnsymbol{footnote}}

\thispagestyle{empty}

\title{Signature of paraparticles: \\ a minimal Gedankenexperiment}
\author{ Francesco Toppan\thanks{{E-mail: {\it toppan@cbpf.br}}}
\\
\\
}
\maketitle

{\centerline{
{\it CBPF, Rua Dr. Xavier Sigaud 150, Urca,}}\centerline{\it{
cep 22290-180, Rio de Janeiro (RJ), Brazil.}}
~\\
\maketitle

\begin{abstract}
~\par
Paraparticles beyond bosons and fermions can be exchanged via either the braid group (anyons, existing up to $D=2$ space dimensions) or the permutation group; in the latter case the space dimensions are not limited. Besides being predicted, anyons have been experimentally detected. The situation differs for paraparticles exchanged via the permutation group (``permutation-group parastatistics").\par The first test to detect their theoretical signature
was published in 2021 (for \zzg parafermions; it was soon followed  by a second paper proving the detectability of \zzg parabosons). Later on, two further papers
proved theoretical signatures of permutation-group parastatistics.
These works demonstrate that, in certain situations, a long-held belief on the  ``conventionality of parastatistics" argument can be evaded: some measurements of permutation-group paraparticles cannot be recovered from ordinary bosons/fermions.\par
The main question now is how to experimentally detect or engineer in the laboratory such paraparticles. For this aim a minimal setup for the theoretical test is here provided: a Gedankenexperiment (a simplified version of the two tests published in 2021) which, essentially, is a {\it flow chart} of logical operations. The key point is to present, to experimentalists, the necessary steps to be simulated/realized in the laboratory (possibly, by manipulating qudits).\par
In this minimal setup, the detection/engineering of paraparticles is mapped into a {\it chirality test}. The mathematical setting is based on \zzg color Lie (super)algebras and derived mathematical structures.
\end{abstract}
\vfill
\rightline{CBPF-NF-001/26~~~~~~~}
\newpage

\section{Introduction}

It is known since 1953 \cite{gre} that paraparticles beyond bosons/fermions and exchanged under the permutation group can be consistently formulated in any space dimension.  On the other hand, a long-held widespread belief (the so-called, see \cite{conventionality},  ``conventionality of parastatistics"  argument based on various locality principles, the most compelling ones given in \cite{{drharo},{dhr},{doro}}) assumed that all conceivable experimental results for these paraparticles could be reproduced by ordinary bosons/fermions statistics. This means that, if the argument is valid and applicable, no distinct signature of parastatistics could be found.\par
I presented, in the  \cite{top1} paper published in 2021, the first theoretical test which proves that paraparticles, exchanged under the permutation group,  admit a distinct signature of parastatistics. The paraparticles under consideration were  Rittenberg-Wyler \cite{{rw1},{rw2}} ${\mathbb Z}_2\times{\mathbb Z}_2$-graded parafermions; this result was soon extended  in \cite{top2} to prove the theoretical detectability of Rittenberg-Wyler ${\mathbb Z}_2\times{\mathbb Z}_2$-graded parabosons.  Later on, almost simultaneously, further proofs of the theoretical detectability of paraparticles exchanged under the permutation group were presented in  \cite{nbits} and \cite{waha}. \par
These four papers \cite{{top1},{top2},{nbits},{waha}} have a common feature: in all discussed cases the locality hypotheses at the basis of the conventionality's argument are (via different mechanisms) evaded.
A new awareness follows, namely, the recognition that  certain classes of quantum models admit a theoretical signature of their parastatistics. Having recognized the theoretical detectability of these paraparticles, the ball is now in the experiment's court. So far, no experiment has yet detected/engineered in the laboratory their presence. As a matter of fact, no such a test has ever been conducted. The reason is simple; an experiment should be designed 
for a system presenting a distinct signature of parastatistics. Here, the \cite{{top1},{top2},{nbits},{waha}} theoretical models come in handy: they could lead, if properly translated into conceivable experiments, to a detection of paraparticles. \par
The aim of this paper is precisely to provide a bridge between theoretical models and conceivable experiments, presenting a Gedankenexperiment which, essentially, consists of a flow chart of logical operations performed in the multi-particle sector of a quantum model.  The main idea is to motivate and make aware of this challenge those experimentalists who have means to implement logical operations into some material support (whatever it is). For this purpose, a minimal signature of parastatistics is quite apt since it is conceivable to be more ubiquous and more easily transcribed into devices. The minimal signature, described in \cite{top1} and \cite{top2}, allows to map the detectability of paraparticles into a chirality-type test. The Gedankenexperiment here presented is a conveniently simplified version of these \cite{{top1},{top2}} tests, with the needed logical operations clearly detailed.
~\par The structure of the paper is the following:\\ 
- Section {\bf 2} gives a very brief account on the state of the art of parastatistics; it is intended to present the conceptual frame behind the proposed Gedankenexperiment;\\
- Section {\bf 3} presents the necessary mathematical background. It introduces the definition of ${\mathbb Z}_2^n$-graded Rittenberg-Wyler color Lie (super)algebras and of their graded Hopf algebras, which are applied to the multi-particle quantization of models; {\it minimal} ${\mathbb Z}_2^2$-graded color Lie (super)algebras and their $4\times 4$ matrix representations are presented.\\
- Section {\bf 4} introduces a minimal, single-particle, ${\mathbb Z}_2^2$-graded, quantum model.\\
- Section {\bf 5}  presents the inequivalent, multi-particle quantizations of this minimal model. They encode the signatures of the respective (para)statistics. A chirality test (the minimal Gedankenexperiment) discriminates ${\mathbb Z}_2^2$-graded paraparticles from ordinary particles.\\
- Section {\bf 6} details the comparative outcomes of the Gedankenexperiment and discusses the challenges for its experimental implementation; \\
- In the Conclusions further perspectives and some related recent advancements are outlined.

\section{A brief state of the art}

A brief state of the art concerning parastatistics (for a more complete account, see \cite{ijgmmp}) is presented. Parastatistics, based on the so-called trilinear relations, were introduced by Green in 1953  \cite{gre}
(parabosons {\it or} parafermions were accommodated in the scheme). In \cite{grme} trilinear relations were extended to theories accommodating both parabosons {\it and} parafermions. \par
The 1953 introduction of parastatistics prompted a natural question: {\it why only bosons and fermions are observed and not the most general class of paraparticles?} To make a long debate short (see \cite{conventionality} for a detailed description) a {\it conventionality of parastatistics' argument} gained traction. Basically, various locality principles point out that no signature of paraparticles should be found since their physics could always be recovered from ordinary bosons/fermions statistics. \par
The conventionality's argument became widely accepted despite (or thanks to!) the fact that its applicability was not challenged. This unquestioned acceptance of the conventionality's argument (also known, see \cite{conventionality}, as the {\it equivalence thesis}) brought consequences. One example for all; for many years the color Lie (super)algebras introduced by Rittenberg-Wyler in 1978 \cite{{rw1},{rw2}} were considered not relevant to physics, since they imply the presence of paraparticles beyond bosons/fermions. Important mathematical investigations of their parastatistics were conducted \cite{{yaji1},{yaji2},{kaha},{kan},{tol2},{stvdj}}, but their physical significance was  questioned under the conventionality's paradigm. \par
The \cite{{rw1},{rw2}} color Lie (super)algebras were introduced as extensions of ordinary Lie (super)algebras; they are graded by arbitrary abelian groups and their graded brackets satisfy generalized Jacobi identities. The ${\mathbb Z}_2^n$ grading groups induce, see \cite{nbits}, the so-called $n$-bit parastatistics. Color Lie algebras admit only parabosons, while color Lie superalgebras also admit parafermions which satisfy \cite{brqm} a generalized Pauli's exclusion principle (for ${\mathbb Z}_2^n$ gradings, the ordinary Pauli's exclusion principle is satisfied). The standard  \cite{kac} Lie superalgebras are recovered as  $n=1$ color Lie superalgebras. For $n\geq 2$, the defining brackets of  ${\mathbb Z}_2^n$-graded color Lie (super)algebras are given  by commutators/anticommutators which are differently organized with respect to ordinary bosons/fermions. 
The $n=2$ ($2$-bit) case introduces the simplest examples of parastatistics.\par
In \cite{brdu} Bruce and Duplij introduced a quantum mechanical model (it belongs to a more general class of classical models \cite{akt1} which can be quantized \cite{akt2}) which, at the same time, is an example of supersymmetric quantum mechanics, but it also possesses a \zzg Lie superalgebra symmetry. The problem of whether the \zzg  symmetry of this model has testable physical consequences was addressed in \cite{top1}. Obviously, for the original, {\it single-particle} Bruce-Duplij quantum model the  \zzg symmetry gives an alternative, but equivalent, description of the system. 
The main question to be answered was whether, in the First-Quantized, multi-particle extension of the model, the \zzg parastatistics could play a role. This question could be reduced to a simple test, based on a combinatorics, with a clear yes/no answer.  The test was positive: the \zzg parafermions, induced by the multi-particle version of the model, are detectable (in \cite{top2} the test was extended to \zzg parabosons). \par Heavily hypothesis-laden philosophical arguments, when conveniently simplified and formalized, can be put to test in a manageable environment; the outcome of the test, either positive or negative, is not predicted in advance. \par
Further developments, concerning theoretical signatures of parastatistics, were obtained in \cite{nbits} and \cite{waha}. It was shown in \cite{nbits}  that signature of paraparticles could be directly read,
for certain multi-particle quantum models, from the degeneracies of their discrete energy eigenvalues; the models under consideration in that work are deformed oscillators whose energy levels are determined by ${\mathbb Z}_2^n$-graded spectrum-generating Lie (super)algebras. The \cite{waha} paper introduced a different approach, presenting detectability tests for quantum spin Hamiltonians. In \cite{waha} four different classes of parastatistics were considered,  the first class being given by \zzg parafermions.\par
The \cite{{top1},{top2},{nbits},{waha}}  papers show that the locality principles underlying the equivalence thesis can be evaded. \cite{{top1},{top2},{nbits}} present a First Quantized formulation where the notion of locality is not required; in \cite{waha}, the locality principle is evaded  due to the excited states being created by non-local operators  of stringy nature.\par
The detectability tests presented in \cite{{top1},{top2},{nbits}} seem to be more directly applicable to paraparticles engineered in the laboratory via quantum information technology. The detectability tests in \cite{waha}, on the other hand, seem suitably applicable to possible emergent quasiparticles in condensed matter physics. A somehow related result \cite{beyond} shows that \zzg parafermionic Hamiltonians, described by \zzg superdivision algebras, go beyond the ``10-fold way", i.e. the \cite{10fold} periodic table of topological insulators and superconductors which accommodates ordinary particles. \par
 It is worth comparing the present situation for this type of parastatistics with that of anyons.
The possibility in low space dimensions of a parastatistics based on the more general braid group, instead of the permutation group recovered as a particular case, was first pointed out in \cite{lemy}.  In \cite{wil}, these paraparticles were named {\it anyons}. For an account of the relevant works leading to their discovery and the following applications one can consult \cite{gol}. Unlike ``permutation-group paraparticles", anyons were never controversial; indeed, as emergent quasiparticles exchanged via the braid group, they do not fall into the equivalence thesis discussed in \cite{conventionality}.\par
Besides being theoretically predicted, quite recently anyons have been experimentally detected. Significant experimental steps are presented in\cite{expanyons} (the first experimental evidence of anyons),  \cite{nonabanyons} 
(the first detection of non-abelian anyons transforming under higher dimensional representations 
of the braid group) and \cite{1danyons} (the first experimental evidence of one-dimensional anyons). At least for anyons, the field of parastatistics reached its maturity both theoretically and experimentally. 

\section{Mathematical preliminaries on ${\mathbb Z}_2^n$-graded,  $n$-bit parastatistics}

The mathematical structures discussed in the paper are, for selfconsistency,  here summarized.
They are: the Rittenberg-Wyler \cite{{rw1},{rw2}} ${\mathbb Z}_2^n$-graded color Lie (super)algebras (where ${\mathbb Z}_2^n:= {\mathbb Z}_2\times \ldots \times {\mathbb Z}_2$ denotes the product of $n$ groups ${\mathbb Z}_2$), the induced graded Hopf algebras \cite{maj} applied to the construction of multi-particle quantum systems and the minimal, see \cite{kuto}, \zzg color Lie (super)algebras with their $4\times 4$ matrix representations. \\
Updated results about color Lie (super)algebras graded by general abelian groups are found in the recent  paper \cite{rbzha}, which is based on the Scheunert's approach \cite{sch} to this class of algebras.

\subsection{The \zzg color Lie (super)algebras}

The Rittenberg-Wyler \cite{{rw1},{rw2}} \zzg color Lie (super)algebras are here introduced following the $2$-bit
(which is easily extended, for ${\mathbb Z}_2^n$ grading abelian groups, to the $n$-bit cases) presentation given in \cite{nbits}. \par
At first one starts with a ${\mathbb Z}_2^n$-graded associative ring of operators $A,B,C,\ldots$ with respective $n$-bit gradings $[A]=\alpha,~[B]=\beta,~[C]=\gamma,~\ldots$.
For $n=1$ the grading bits are $\{0,1\}$, for $n=2$ they are
$\{00,10,01,11\}$ and so on. The grading is consistent with the standard addition and multiplication of operators. In particular, the multiplication satisfies
\bea
[A\cdot B] = \alpha+\beta &&{\textrm{with $mod~2$ addition}}.
\eea
The grading  $[{\mathbb I}]:={\underline{0}}$ of the identity operator ${\mathbb I}$ is  the zero element satisfying
${\underline{0}}+\alpha=\alpha+{\underline{0}}=\alpha$ for any $\alpha$ (for $n=2$, the ${\underline{0}} $ grading is identified with the $2$-letter word $00$).\par
~\par
For the ${\mathbb Z}_2^n$ abelian groups, the inequivalent \cite{{rw1},{rw2},{sch}} consistency conditions for the introduction of color Lie (super)algebras  can be stated, see \cite{nbits}, as follows:\\
~\\
{\it i}) a $\langle ,\rangle $ bilinear mapping is introduced, given by
\bea\label{bilinear}  \langle ,\rangle &:& {\mathbb Z}_2^n\times{\mathbb Z}_2^n\rightarrow {\mathbb Z}_2.
\eea
Without loss of generality it can be assumed to be symmetric:
$\langle\beta,\alpha\rangle = \langle\alpha,\beta\rangle\in \{0,1\}\quad \forall \alpha,\beta$;\\
{\it ii}) the further condition, which implies a graded Leibniz rule, is imposed:
\bea\label{condition}
 \langle \alpha , \beta+\gamma\rangle &=& \langle \alpha,\beta\rangle +\langle \alpha,\gamma\rangle {\textrm{~~ mod~~$2$}}.
\eea
\\\
~\\
A graded Lie bracket $(,)$ is introduced through the position
\bea\label{gradedbracket}
(A,B)&:=&A\cdot B -(-1)^{\langle\alpha,\beta\rangle} B\cdot A.
\eea

The right hand side is either a commutator (for ${\langle\alpha,\beta\rangle} =0$) or an anticommutator (for ${\langle\alpha,\beta\rangle} =1$).
A set of $A,B,C,\ldots $ generators, closed under the (\ref{gradedbracket}) graded brackets, defines a color Lie (super)algebra satisfying the graded Jacobi identity
\bea\label{jacobi}
(-1)^{\langle \gamma,\alpha\rangle} (A,(B,C))+(-1)^{\langle \alpha,\beta\rangle} (B,(C,A))+(-1)^{\langle \beta,\gamma\rangle} (C,(A,B))&=&0.
\eea
Following Rittenberg-Wyler's definition, a ${\mathbb Z}_2^n$-graded color Lie algebra is obtained if, for any $\alpha$ entering (\ref{bilinear}), the $ \langle\alpha ,\alpha \rangle=0$ condition is satisfied. Otherwise, if there exists at least a ${\overline \alpha}\in{\mathbb Z}_2^n$ such that $ \langle{\overline \alpha} ,{\overline \alpha} \rangle=1$, one gets
a ${\mathbb Z}_2^n$-graded color Lie superalgebra. The physical implications of this difference are discussed in Section {\bf 5}.\par
~\par
For $n=2$ there are four classes of ${\mathbb Z}_2^2$-graded color Lie (super)algebras induced by the 
inequivalent symmetric maps (\ref{bilinear}) satisfying the (\ref{condition}) condition. Each one of this four classes can be presented as a $4\times 4$ array. The rows/columns are respectively labeled by the $2$-bit gradings $\alpha$ and $\beta$, while the corresponding $0,1$ entry denotes the $\langle\alpha,\beta\rangle$ value. \par
The four admissible cases, respectively labeled as ``$LA$", ``$LS$", ``$CLA$", ``$CLS$", are:
\bea \label{n2cases}
{\textrm{I - $``LA"$:}} ~~~&&\begin{array}{c|cccc}&00&10&01&11\\ \hline 00&0&0&0&0\\10&0&0&0&0\\01&0&0&0&0\\11&0&0&0&0
\end{array}, ~~ {\textrm{producing an ordinary Lie Algebra;}}
\nonumber\\
&&\nonumber\\
{\textrm{II - $``LS"$:}} ~~&& 
\begin{array}{c|cccc}&00&10&01&11\\ \hline 00&0&0&0&0\\10&0&1&1&0\\01&0&1&1&0\\11&0&0&0&0
\end{array},~~
 {\textrm{producing an ordinary Lie Superalgebra;}}
\nonumber\\&&\nonumber\\
{\textrm{III - $``CLA"$:}} ~&& 
\begin{array}{c|cccc}&00&10&01&11\\ \hline 00&0&0&0&0\\10&0&0&1&1\\01&0&1&0&1\\11&0&1&1&0
\end{array},~~
 {\textrm{giving a nontrivial ${\mathbb Z}_2^2$-graded Color Lie Algebra;}}\nonumber\\
&&\nonumber\\
{\textrm{IV - $``CLS"$:}}&& 
\begin{array}{c|cccc}&00&10&01&11\\ \hline 00&0&0&0&0\\10&0&1&0&1\\01&0&0&1&1\\11&0&1&1&0
\end{array},~~
 {\textrm{giving a nontrivial ${\mathbb Z}_2^2$-graded Color Lie Superalgebra.}}\nonumber\\&&
\eea
The colored cases III and IV imply the existence of a parastatistics. 

\subsection{Graded Hopf algebras with braided tensor product}

Graded Hopf algebras endowed with a braided tensor product allow to introduce, in a First-Quantized formalism,  the multi-particle sectors of a quantum model \cite{maj}. \par
Given a color Lie (super)algebra ${\mathfrak{g}}$ graded by an abelian group one can introduce, see \cite{rbzha}, the color Universal Enveloping algebra $U:={\cal U} ({\mathfrak g})$ which has the structure of a color Hopf algebra. It possesses, in particular, a coproduct map $\Delta$ which is the relevant operation allowing to construct multiparticle states:
\bea\label{coproduct}
\Delta &:& U\rightarrow U\otimes U.
\eea 
The coproduct satisfies the coassociativity property
\bea\label{coassoc}
 &\Delta^{m+1}:=  (\Delta\otimes{\mathbf 1})\Delta^m=({\mathbf 1}\otimes \Delta)\Delta^m \qquad ~ {\textrm{(where~ $\Delta^1\equiv \Delta$)}}
\eea
and comultiplication
\bea\label{comult}
\Delta(u_1u_2) &=& \Delta(u_1)\cdot \Delta(u_2) \qquad {\textrm{for any ~$u_1,u_2\in U$.}}
\eea

Let us now specialize the following discussion to the  ${\mathbb Z}_2^n$ grading abelian groups. The tensor product ``$\otimes$" entering (\ref{coproduct}) is {\it braided}. This means that, for $A,B,C,D\in {\cal U} ({\mathfrak g}) $ with respective $\alpha,\beta,\gamma,\delta$ gradings, it satisfies the relation
\bea\label{braidedtensor}
(A\otimes B)\cdot (C\otimes D) &=& (-1)^{\langle\beta,\gamma\rangle}(AC)\otimes (BD).
\eea
${\langle\beta,\gamma\rangle}$ in the right hand side denotes the symmetric bilinear map introduced in (\ref{bilinear}). \par

The action of the coproduct on the identity ${\bf 1}\in {\cal U}({\mathfrak{g}})$ and on the primitive elements $g\in{\mathfrak{g}}$ is
\bea\label{coproductaction}
\Delta({\bf 1})={\bf 1}\otimes{\bf 1}, \quad&&\quad
\Delta(g) = {\bf 1}\otimes g+g\otimes {\bf 1}.
\eea 
It follows that $\Delta(u)\in U\otimes U$ is recovered, for a generic $u\in U\equiv {\cal U}({\mathfrak{g}})$, from  (\ref{coproductaction}) and (\ref{comult}).\par
~\\
In physical applications, Hamiltonians and creation/annihilation operators are typical examples of primitive elements. The coproduct $\Delta=\Delta^1$ is used in the construction of $2$-particle states ($\Delta^m$ from (\ref{coassoc}) is employed in the construction of $(m+1)$-particle states). 

\subsection{Minimal ${\mathbb Z}_2^2$-graded color Lie (super)algebras}

The notion of {\it minimal}, ${\mathbb Z}_2^2$-graded color Lie algebras and superalgebras was introduced in \cite{kuto}. They are here respectively denoted as ``${\mathfrak{ min}}_A$" and ``${\mathfrak{ min}}_S$".  For $\star\equiv A, S $ a minimal (super)algebra possesses
the (\ref{gradedbracket}) graded bracket satisfying the (\ref{jacobi}) graded Jacobi identity. Denoting as $(,)_\star$ the respective graded bracket one has $(,)_\star: {\mathfrak{ min}}_\star\times{\mathfrak{ min}}_\star\rightarrow {\mathfrak{ min}}_\star$. The suffix $A,S$ indicates if the graded bracket induced by the arrays presented in  (\ref{n2cases}) corresponds to either the III case  (with ``A" standing for a nontrivial color Lie algebra) or the IV case
(with ``S" standing for a nontrivial color Lie superalgebra). The minimality condition further requires ${\mathfrak{ min}}_\star$ being spanned by one and only one generator belonging to each graded sector $00,10,01,11$ (for a total number of $4$  generators spanning  ${\mathfrak{ min}}_\star$). The classification of the inequivalent ${\mathfrak{ min}}_A$ color Lie algebras and ${\mathfrak{ min}}_S$ color Lie superalgebras was presented in \cite{kuto}.
\par
The minimal representations of ${\mathbb Z}_2\times {\mathbb Z}_2$-graded color Lie (super)algebras are given by $4\times 4$ matrices; the nonvanishing entries of their respective graded sectors, expressed by the symbol ``$\ast$", are:
{\small{\bea\label{gradedmatrices}
M_{00}=\left(\begin{array}{cccc} \ast&0&0&0\\0&\ast&0&0\\0&0&\ast&0\\0&0&0&\ast\end{array}\right)\in {\cal G}_{00},&&M_{10}=\left(\begin{array}{cccc} 0&\ast&0&0\\\ast&0&0&0\\0&0&0&\ast\\0&0&\ast&0\end{array}\right)\in {\cal G}_{10},\nonumber\\
M_{01}=\left(\begin{array}{cccc} 0&0&\ast&0\\0&0&0&\ast\\\ast&0&0&0\\0&\ast&0&0\end{array}\right)\in {\cal G}_{01},&&M_{11}=\left(\begin{array}{cccc} 0&0&0&\ast\\0&0&\ast&0\\0&\ast&0&0\\\ast&0&0&0\end{array}\right)\in {\cal G}_{11}.
\eea}}
In physical applications the entries are either $c$-numbers or differential operators.\\
The algebra of the general $4\times 4$ graded matrices (\ref{gradedmatrices}), closed under addition and multiplication, is denoted as ${\mathfrak{gl}}(1|1|1|1)$.\par
The matrix product of two $4\times 4$, ${\mathbb Z}_2^2$-graded, matrices $M',M''$ produces a graded matrix $M$:
\bea
\quad M'_{i'j'}\cdot M''_{i''j''} &=& M_{ij}, \qquad {\textrm{with ~ $i=i'+i''$~ and ~ $j=j'+j''$ ~ (mod $2$).}}
\eea
Tensoring the two matrices $M',M''$  produces a $16\times 16$, ${\mathbb Z}_2^2$-graded, matrix $M^{(2)}$:
\bea
\quad M'_{i'j'}\otimes M''_{i''j''} &=& M^{(2)}_{ij}, \qquad {\textrm{with ~ $i=i'+i''$ and ~ $j=j'+j''$ ~ (mod $2$).}}
\eea

As discussed in Section {\bf 5},  a $2$-particle observable belongs to the $00$-graded sector; its non-vanishing entries belong to the following $16\times 16$, ${\mathbb Z}_2^2$-graded, matrix:
\bea\label{2particleobs}
M_{00}^{(2)}&=&\left(\begin{array}{cccccccc|cccccccc}
\ast&0&0&0&0&\ast&0&0&0&0&\ast&0&0&0&0&\ast\\
0&\ast&0&0&\ast&0&0&0&0&0&0&\ast&0&0&\ast&0\\
0&0&\ast&0&0&0&0&\ast&\ast&0&0&0&0&\ast&0&0\\
0&0&0&\ast&0&0&\ast&0&0&\ast&0&0&\ast&0&0&0\\
0&\ast&0&0&\ast&0&0&0&0&0&0&\ast&0&0&\ast&0\\
\ast&0&0&0&0&\ast&0&0&0&0&\ast&0&0&0&0&\ast\\
0&0&0&\ast&0&0&\ast&0&0&\ast&0&0&\ast&0&0&0\\
0&0&\ast&0&0&0&0&\ast&\ast&0&0&0&0&\ast&0&0\\ \hline
0&0&\ast&0&0&0&0&\ast&\ast&0&0&0&0&\ast&0&0\\
0&0&0&\ast&0&0&\ast&0&0&\ast&0&0&\ast&0&0&0\\
\ast&0&0&0&0&\ast&0&0&0&0&\ast&0&0&0&0&\ast\\
0&\ast&0&0&\ast&0&0&0&0&0&0&\ast&0&0&\ast&0\\
0&0&0&\ast&0&0&\ast&0&0&\ast&0&0&\ast&0&0&0\\
0&0&\ast&0&0&0&0&\ast&\ast&0&0&0&0&\ast&0&0\\
0&\ast&0&0&\ast&0&0&0&0&0&0&\ast&0&0&\ast&0\\
\ast&0&0&0&0&\ast&0&0&0&0&\ast&0&0&0&0&\ast
\end{array}\right).
\eea
~\\
In terms of the minimal $4\times 4$ matrix representations, the minimal \zzg color Lie (super)algebras ${\mathfrak{ min}}_A, {\mathfrak{ min}}_S$ are color subalgebras of the general, $4\times 4$, \zzg linear matrices ${\mathfrak{gl}}(1|1|1|1)$. For minimal color Lie (super)algebras defined by the respective graded brackets one has:
 \bea
&&{\textrm{for color Lie algebras, corresponding to the III case in (\ref{n2cases}):}}\qquad  ~~~~~ {\mathfrak{ min}}_A\subset  {\mathfrak{gl}}_{A}(1|1|1|1);   \nonumber\\
&&{\textrm{for color Lie superalgebras, corresponding to the IV case in (\ref{n2cases}):}} ~~~~ {\mathfrak{ min}}_S \subset {\mathfrak{gl}}_{S}(1|1|1|1).\nonumber
\eea
The $A,S\equiv \star$ suffix in ${\mathfrak{gl}}_{\star}(1|1|1|1)$ indicates which graded brackets are defined as\\
(anti)commutators for the  ${\mathfrak{gl}}(1|1|1|1)$ matrix generators.

\section{The minimal, {single-particle}, ${\mathbb Z}_2\times{\mathbb Z}_2$-graded quantum model}

As already mentioned, the minimal scenario presenting a signature of {\it permutation-group parastatistics} is based on paraparticles accommodated in  a \cite{{rw1},{rw2}} Rittenberg-Wyler \zzg color Lie algebra or color Lie superalgebra.  Two different approaches to parastatistics have been applied to color Lie (super)algebras; a formulation via Green's \cite{gre} trilinear relations was discussed in  \cite{{yaji1},{yaji2},{kaha},{kan},{tol2},{stvdj}}, while  the Majid's \cite{maj}  approach based on graded Hopf algebras was presented in \cite{{kan},{top1},{top2},{nbits}}. In the literature, the connection between these two approaches to parastatistics was clarified in \cite{{anpo},{kada}}. The Majid's approach controls the construction of the {\it multi-particle} sectors in a First-Quantized formulation of a quantum model.  Following \cite{{top1},{top2}}, this approach is applied to the Gedankenexperiment here presented. \par
The {\it minimality} of the proposed Gedankenexperiment refers to two distinct features:\\
{\it i})  the \zzg ($2$-bit) parastatistics is the minimal setting to introduce paraparticles and\\
{\it ii}) the quantum model under consideration is given by the simplest possible Hamiltonian fitting in this framework; it belongs, in the {\it single-particle} sector, to the diagonal part of a ${\mathbb Z}_2\times{\mathbb Z}_2$-graded, $4\times 4$ matrix.  The ${\mathbb Z}_2\times{\mathbb Z}_2$-graded, spectrum-generating, color Lie (super)algebras of the model are minimal (super)algebras in the sense specified in \cite{kuto} (that is, they possess one and only one generator in each graded sector).\par
This Section presents the formulation of the  minimal, {\it single-particle}, ${\mathbb Z}_2\times{\mathbb Z}_2$-graded quantum model and of its spectrum-generating graded (super)algebras. 

\subsection{The quantum model}\par

The minimal quantum Hamiltonian $H_{min}$ is given by a $4\times 4$ diagonal matrix  such that \\$H_{min}= diag(0,\lambda_1,\lambda_2,\lambda_3=\lambda_1+\lambda_2)$ where, without loss of generality, the vacuum energy is set to zero. To simplify the analysis the energy levels are assumed not to be degenerate. The real eigenvalues are set to be $\lambda_3>\lambda_2>\lambda_1>0$. The energy of the first excited state is conveniently normalized to be $\lambda_1=1$.
We can therefore set $\lambda_2:=\lambda$ and $\lambda_3 =\lambda+1$. With these positions $H_{min}$ reads

{\footnotesize{\bea
H_{min}&=& \left(\begin{array}{cccc} 0&0&0&0\\0&1&0&0\\0&0&\lambda&0\\0&0&0&\lambda+1\end{array}\right), \qquad
{\textrm{for ~ $\lambda>1$}}.
\eea
}}

The Hamiltonian $H_{min}$ belongs to the $00$-graded sector of the ${\mathbb Z}_2^2$-graded matrices (\ref{gradedmatrices}). \\
The energy level $\lambda$ is assumed to be an externally controlled parameter (for instance, it could be related to an external, constant, electric or magnetic field).\par
In \cite{{top1},{top2}} the constant Hamiltonian $H_{min}$ enters a $4\times 4$ matrix oscillator $H_{osc}'$ defined as 
\bea
H_{osc}'&=& H_{osc}+ H_{min}, \quad~~ {\textrm{where~~ $H_{osc}=\frac{1}{2}(\partial_x^2+x^2)\cdot {\mathbb I}_4$}}
\eea
(here and throughout the text, the ``${\mathbb I}_n$" symbol denotes the $n\times n$ identity matrix).
For the present purposes it is not needed to extend the analysis to the oscillator Hamiltonian $H_{osc}'$.

\subsection{Equivalent, single-particle, spectrum-generating algebras}

The model under consideration admits, as spectrum-generating algebras, either a pair of fermionic oscillators or  a pair  of ${\mathbb Z}_2^2$-graded parafermionic oscillators;  in both cases the two pairs of oscillators allow to construct the excited energy states of the model. These two pairs of oscillators are introduced as follows.\par

At first we define the $2\times 2$ matrices
{\footnotesize{\bea
&& \beta=\left(\begin{array}{cc} 0&1\\0&0\end{array}\right),\qquad \gamma=\left(\begin{array}{cc} 0&0\\1&0\end{array}\right),\qquad X=\left(\begin{array}{cc} 1&0\\0&-1\end{array}\right), \qquad I\equiv {\mathbb I}_2= 
\left(\begin{array}{cc} 1&0\\0&1\end{array}\right).\eea}}
The pair of fermionic oscillators is realized through the positions:
\bea
f_1 =\beta\otimes I,\qquad  f_1^\dagger=\gamma\otimes I,\qquad  f_2 =X\otimes \beta,\qquad  f_2^\dagger =X\otimes\gamma, \qquad c=I\otimes I. &&
\eea
The above $5$ operators are the generators of the  ${\mathfrak{h}}_{fer}(2)$ Heisenberg-Lie superalgebra (where  $f_1,f_1^\dagger,f_2,f_2^\dagger, c\in {\mathfrak{h}}_{fer}(2)$),  which is defined by the (anti)commutators
\bea\label{hfer}
&&\{f_1,f_1\}=\{f_2,f_2\}=\{f_1^\dagger,f_1^\dagger\}=\{f_2^\dagger,f_2^\dagger\}=0,\nonumber\\
&&\{f_1,f_1^\dagger\}=\{f_2,f_2^\dagger\}=c,\qquad \quad [c,z] = 0 \quad \quad \forall z\in {\mathfrak{h}}_{fer}(2),\nonumber\\
&&\{f_1,f_2\}=\{f_1,f_2^\dagger\}=\{f_1^\dagger,f_2\}=\{f_1^\dagger,f_2^\dagger\}=0.
\eea

The pair of ${\mathbb Z}_2^2$-graded parafermionic oscillators is realized through the positions:
\bea
p_1 =\beta\otimes I,\qquad  p_1^\dagger=\gamma\otimes I,\qquad  p_2 =I\otimes \beta,\qquad  p_2^\dagger =I\otimes\gamma, \qquad c=I\otimes I. &&
\eea
The above $5$ operators are the generators of the  ${\mathbb Z}_2^2$-graded color ${\mathfrak{h}}_{pf}(2)$ Heisenberg-Lie superalgebra (with $p_1,p_1^\dagger,p_2,p_2^\dagger, c\in {\mathfrak{h}}_{pf}(2)$), which is defined by the (anti)commutators
\bea\label{hpf}
&&\{p_1,p_1\}=\{p_2,p_2\}=\{p_1^\dagger,p_1^\dagger\}=\{p_2^\dagger,p_2^\dagger\}=0,\nonumber\\
&&\{p_1,p_1^\dagger\}=\{p_2,p_2^\dagger\}=c,\qquad \quad [c,z] = 0 \quad \quad \forall z\in {\mathfrak{h}}_{pf}(2),\nonumber\\
&&[p_1,p_2]=[p_1,p_2^\dagger]=[p_1^\dagger,p_2]=[p_1^\dagger,p_2^\dagger]=0.
\eea
The difference of (\ref{hpf}) with respect to (\ref{hfer}) is in the third line, given by commutators instead of anticommutators.\par
This difference can be encoded in a $\varepsilon=\pm 1$ sign, with $\varepsilon =-1$ for fermions and $\varepsilon =+1$ for parafermions. Indeed, we can set
{\footnotesize{\bea
{\overline f}_{1} := \left(\begin{array}{cccc}
0&0&1&0\\
0&0&0&1\\
0&0&0&0\\
0&0&0&0
\end{array}\right)= f_1=p_1,&& 
{\overline f}_{1}^\dagger := \left(\begin{array}{cccc}
0&0&0&0\\
0&0&0&0\\
1&0&0&0\\
0&1&0&0
\end{array}\right)= f_1^\dagger=p_1^\dagger
\eea
}}
and
{\footnotesize{\bea
&{\overline f}_{2,\varepsilon} := \left(\begin{array}{cccc}
0&1&0&0\\
0&0&0&0\\
0&0&0&\varepsilon\\
0&0&0&0
\end{array}\right),\qquad 
{\overline f}_{2,\varepsilon}^\dagger := \left(\begin{array}{cccc}
0&0&0&0\\
1&0&0&0\\
0&0&0&0\\
0&0&\varepsilon&0
\end{array}\right), \qquad c = \left(\begin{array}{cccc}
1&0&0&0\\
0&1&0&0\\
0&0&1&0\\
0&0&0&1
\end{array}\right),&
\eea
}}
so that
{{\bea
&{\overline f}_{2,\varepsilon=-1} = f_2,  \qquad ~~
{\overline f}_{2,\varepsilon=-1}^\dagger = f_2^\dagger, \qquad  \qquad
{\overline f}_{2,\varepsilon=+1} = p_2,  \qquad ~~
{\overline f}_{2,\varepsilon=+1}^\dagger = p_2^\dagger.
&
\eea}}
One should note that these matrices belong to the following ${\mathbb Z}_2^2$-graded sectors of (\ref{gradedmatrices}):
\bea
&{\overline f}_{1}, ~
{\overline f}_{1}^\dagger ~\in~ {\cal G}_{10}, \qquad \qquad
{\overline f}_{2,\varepsilon}, ~
{\overline f}_{2,\varepsilon}^\dagger ~\in ~{\cal G}_{01}, \qquad \qquad c~\in~{\cal G}_{00}.&
\eea

The ordinary Lie superalgebra (\ref{hfer}) corresponds to the (anti)commutators  given by the II case of (\ref{n2cases}), while the  color Lie superalgebra (\ref{hpf}) corresponds to the (anti)commutators given by the IV case of (\ref{n2cases}).\par
~\par
The  minimal Hamiltonian $H_{min}$ can be described either as
\bea
H_{min} &=& f_1^\dagger f_1 +\lambda f_2^\dagger  f_2 \qquad{\textrm{(fermionic description)  \qquad or}}\nonumber\\
H_{min} &=& p_1^\dagger p_1 +\lambda p_2^\dagger p_2 \qquad{\textrm{(parafermionic description).}}
\eea
The Hilbert space of the single-particle model is the $4$-dimensional, ${\mathbb Z}_2^2$-graded, vector space $V$ (with $V=V_{00}\oplus V_{10}\oplus V_{01}\oplus V_{11}$), spanned by the vectors $v_i$, for $i=1,2,3,4$:

{\footnotesize{
\bea\label{vspace}
&v_1=\left(\begin{array}{c}
1\\
0\\
0\\
0
\end{array}\right) \in V_{00}, \qquad v_2=\left(\begin{array}{c}
0\\
1\\
0\\
0
\end{array}\right) \in V_{10},\qquad  v_3=\left(\begin{array}{c}
0\\
0\\
1\\
0
\end{array}\right) \in V_{01},\qquad v_4=\left(\begin{array}{c}
0\\
0\\
0\\
1
\end{array}\right) \in V_{11}. &
\eea
}}

They are the normalized energy eigenvectors of $H_{min}$; the spectrum of discrete energy eigenvalues is given by $E=0,1,\lambda,\lambda+1$. One gets
\bea
&&{\overline v}_{0}:= v_1, \qquad\quad~~{\overline v}_{1}:= v_3,\qquad \quad~~ {\overline v}_{\lambda}:= v_2,\qquad\quad~~ {\overline v}_{\lambda+1}:= v_4,\qquad\quad ~~ {\textrm{with}}\nonumber\\
&&H_{min}{\overline  v}_0 = 0,\quad ~ H_{min}{\overline  v}_1 = {\overline v}_1,\quad ~H_{min}{\overline  v}_\lambda = \lambda {\overline v}_\lambda,\quad~ H_{min}{\overline  v}_{\lambda+1} = (\lambda+1) {\overline v}_{\lambda +1}.
\eea
In both descriptions, $|vac\rangle:={\overline v}_0=v_1$ is the Fock vacuum of the model, satisfying
\bea\label{fock}
f_1|vac\rangle=f_2|vac\rangle =0 &\quad{\textrm{and}}\quad & p_1|vac\rangle=p_2|vac\rangle =0.
\eea
The creation operators $f_1^\dagger, f_2^\dagger$  from (\ref{hfer})  create the excited states of the model since
\bea
&[H_{min}, f_1^\dagger] = f_1^\dagger, \quad [H_{min}, f_2^\dagger] = \lambda    f_2^\dagger \quad{\textrm{and}}\quad [H_{min}, f_3^\dagger] = (\lambda+1) f_3^\dagger \quad {\textrm{for \quad $f_3^\dagger:= f_1^\dagger f_2^\dagger$,}}&\nonumber\\
&{\textrm{so that}} \qquad\quad {\overline v}_1 = f_1^\dagger {\overline v}_0, \qquad\quad {\overline v}_\lambda = f_2^\dagger {\overline v}_0, \qquad\quad {\overline v}_{\lambda+1} = f_3^\dagger {\overline v}_0.&
\eea
Similarly, the creation operators $p_1^\dagger, p_2^\dagger$  from (\ref{hpf})  create the excited states since
\bea
&[H_{min}, p_1^\dagger] = p_1^\dagger, \quad [H_{min}, p_2^\dagger] = \lambda    p_2^\dagger \quad{\textrm{and}}\quad [H_{min}, p_3^\dagger] = (\lambda+1) p_3^\dagger \quad {\textrm{for \quad $p_3^\dagger:= p_1^\dagger p_2^\dagger$,}}&\nonumber\\
&{\textrm{so that}} \qquad\quad {\overline v}_1 = p_1^\dagger {\overline v}_0, \qquad\quad {\overline v}_\lambda = p_2^\dagger {\overline v}_0, \qquad\quad {\overline v}_{\lambda+1} = p_3^\dagger {\overline v}_0.&
\eea
The fermionic oscillators (\ref{hfer}) and the parafermionic oscillators (\ref{hpf}) produce two different, but physically equivalent, spectrum-generating algebras for the $H_{min}$ single-particle quantum model.
\par
Let us now connect these two constructions with {\it minimal} spectrum-generating algebras, spanned by a single generator in each graded sector. \par
In the fermionic description we can set the four generators of a minimal spectrum-generating algebra to be
\bea
&H_{min}\in {\cal G}_{00}, \quad f_1^\dagger\in {\cal G}_{10},\quad  f_2^\dagger\in {\cal G}_{01},\quad  f_3^\dagger\in {\cal G}_{11}.&
\eea

In the parafermionic description we can set the four generators to be
\bea
&H_{min}\in {\cal G}_{00}, \quad p_1^\dagger\in {\cal G}_{10},\quad  p_2^\dagger\in {\cal G}_{01},\quad  p_3^\dagger\in {\cal G}_{11}.&
\eea
For these choices of generators we apply the four sets of (anti)commutators listed in (\ref{n2cases}). They correspond to case I (i.e., Lie algebra, denoted as $LA$), case II (Lie superalgebra, denoted as $LS$), case III (${\mathbb Z}_2^2$-graded color Lie algebra, denoted as $CLA$) and case IV (${\mathbb Z}_2^2$-graded color Lie superalgebra, denoted as $CLS$). The results are reported in the following Subsection.

\subsection{The minimal, single-particle, spectrum-generating algebras}

For both fermionic (\ref{hfer}) and parafermionic (\ref{hpf}) descriptions, the {minimal} $LA, LS, CLA, CLS$ spectrum-generating algebras are defined by the respective (anti)commutators. A total number of $8=4+4$ different cases are obtained; the fermionic ones will be denoted as  $fLA_{min}$, $fLS_{min}$, $fCLA_{min}$, $fCLS_{min}$ (the parafermionic ones as  $pLA_{min}$, $pLS_{min}$, $pCLA_{min}$, $pCLS_{min}$). \par
~\par
For fermions, the  commutators involving $H_{min}$ are common to all four cases:
\bea \label{minimalfer0}
&[H_{min}, f_1^\dagger] = f_1^\dagger, ~\quad [H_{min}, f_2^\dagger] = \lambda   f_2^\dagger,\quad  [H_{min}, f_3^\dagger] = (\lambda+1) f_3^\dagger.
\eea 
The remaining (anti)commutators are
\bea\label{minimalfer}
{\textrm{for ~~~$fLA_{min}$:}}&& [f_1^\dagger, f_2^\dagger] = 2f_3^\dagger, \quad [f_1^\dagger,f_3^\dagger]=[f_2^\dagger, f_3^\dagger]=0;\nonumber\\
{\textrm{for ~~~$fLS_{min}$:}}&& \{f_1^\dagger,f_1^\dagger\}=\{f_2^\dagger,f_2^\dagger\} =\{f_1^\dagger, f_2^\dagger\}=0, \quad [f_1^\dagger, f_3^\dagger] = [f_2^\dagger,f_3^\dagger]=0;\nonumber\\
{\textrm{for ~$fCLA_{min}$:}}&& \{f_1^\dagger, f_2^\dagger\} = \{f_2^\dagger,f_3^\dagger\}=\{f_3^\dagger, f_1^\dagger\}=0;\nonumber\\
{\textrm{for ~$fCLS_{min}$:}}&& \{f_1^\dagger,f_1^\dagger\}=\{f_2^\dagger,f_2^\dagger\} =0,  \quad [f_1^\dagger, f_2^\dagger]= 2 f_3^\dagger, \quad \{f_1^\dagger, f_3^\dagger\} = \{f_2^\dagger,f_3^\dagger\}=0.
\eea
~\\
Similarly,  for the (\ref{hpf}) parafermionic description, the commutators involving $H_{min}$, 
\bea \label{minimalpf0}
&[H_{min}, p_1^\dagger] = p_1^\dagger, ~\quad [H_{min}, p_2^\dagger] = \lambda   p_2^\dagger,\quad  [H_{min}, p_3^\dagger] = (\lambda+1) p_3^\dagger,
\eea 
are common to all four cases. The remaining (anti)commutators are
\bea\label{minimalpf}
{\textrm{for ~~~$pLA_{min}$:}}&& [p_1^\dagger, p_2^\dagger] =  [p_2^\dagger, p_3^\dagger]=[p_3^\dagger, p_1^\dagger]=0;\nonumber\\
{\textrm{for ~~~$pLS_{min}$:}}&&\{p_1^\dagger,p_1^\dagger\}=\{p_2^\dagger,p_2^\dagger\} =0,  \quad \{p_1^\dagger, p_2^\dagger\}= 2 p_3^\dagger, \quad [p_1^\dagger, p_3^\dagger] = [p_2^\dagger,p_3^\dagger]=0; \nonumber\\ 
{\textrm{for ~$pCLA_{min}$:}}&& \{p_1^\dagger, p_2^\dagger\} = 2p_3^\dagger, \quad \{p_1^\dagger,p_3^\dagger\}=\{p_2^\dagger, p_3^\dagger\}=0;\nonumber\\
{\textrm{for ~$pCLS_{min}$:}}&& \{p_1^\dagger,p_1^\dagger\}=\{p_2^\dagger,p_2^\dagger\} =0, \quad [p_1^\dagger, p_2^\dagger]=0, \quad \{p_1^\dagger, p_3^\dagger\} = \{p_2^\dagger,p_3^\dagger\}=0.
\eea
The eight minimal algebras from (\ref{minimalfer0},\ref{minimalfer}) and (\ref{minimalpf0},\ref{minimalpf})  are all, physically equivalent, spectrum-generating algebras for the single-particle quantum model.\par
~\par
Four extra, physically equivalent, spectrum-generating algebras are obtained as subalgebras; the fermionic ones are spanned by $H_{min}, f_1^\dagger, f_2^\dagger$, while the parafermionic ones are spanned by $H_{min}, p_1^\dagger, p_2^\dagger$.  They correspond to the cases where the (anti)commutators involving $f_1^\dagger , f_2^\dagger$ (respectively, $p_1^\dagger, p_2^\dagger$) are vanishing, leaving a $3$-generator graded (super)algebra with an empty $11$-graded sector (${\cal G}_{11}=\emptyset$). The four extra cases are given by
\bea\label{sub}
fLS_{sub}\subset fLS_{min}, && 
fCLA_{sub}\subset fCLA_{min}\qquad {\textrm{(spanned by $H_{min}, f_1^\dagger, f_2^\dagger$) \quad and}}\nonumber\\
pLA_{sub}\subset pLA_{min}, && 
pCLS_{sub}\subset pCLS_{min}\qquad~ {\textrm{(spanned by $H_{min}, p_1^\dagger, p_2^\dagger$).}}
\eea

Therefore, we end up with a total number of $12=8+4$,  physically equivalent, spectrum-generating algebras for the single-particle model; the complete list is given by (\ref{minimalfer}, \ref{minimalpf}, \ref{sub}).\par
~\par
In the four, {\it minimal},  {\it colored} cases (that is, $fCLA_{min}, ~pCLA_{min},~ fCLS_{min}, ~pCLS_{min}$) the corresponding  algebras are identified with certain minimal color Lie (super)algebras entering the \cite{kuto} classification. \par
It is easily checked, by rescaling the generators, how to recover $fCLA_{min}$, $pCLA_{min}$ from the Table $1$  of minimal color Lie algebras presented in \cite{kuto}. The algebra $fCLA_{min}$ is identified with
$A8_{y,z}$ for the choice $y=\lambda, z=\lambda+1$ of the parameters, while  $pCLA_{min}$ is identified with the minimal color Lie algebra $A6_x$, with the parameter $x$ set to be $x=\frac{1}{2}\frac{\lambda-1}{\lambda+1}$. \par
It follows that
\bea
fCLA_{min} \equiv A8_{\lambda,\lambda+1},  &\quad& pCLA_{min} \equiv A6_{\frac{1}{2}\frac{\lambda-1}{\lambda+1}}.
\eea
~\\
The color Lie superalgebras $fCLS_{min}, pCLS_{min}$ are recovered from Table $2$ of \cite{kuto}. The superalgebra
$fCLS_{min}$ is identified with $S21_y$ for $y=\lambda$, while $pCLS_{min}$ is identified with $S18_{y,z}$ for the $y=\lambda, ~z=\lambda+1$ choice of the parameters. \par
It follows that
\bea
fCLS \equiv S21_{\lambda},  &\quad& pCLS \equiv S18_{\lambda,\lambda+1}.
\eea

\section{Signature of parastatistics in a multi-particle sector}

The graded, {\it single-particle}, spectrum-generating algebras introduced in the previous Section are extended, following the \cite{maj} Majid's approach, into
{\it multi-particle} spectrum-generating algebras. We will see that, in the multi-particle sectors, they induce inequivalent Hilbert spaces presenting signatures of (para)statistics.  The complete analysis is here provided
for the $2$-particle sector; this sector is sufficient for our scope of discussing the minimal Gedankenexperiment.\par
Before presenting the results we summarize the construction of the spectrum-generating algebras obtained in the previous Section and outline,  following the discussion in Subsection {\bf 3.2}, the use of graded Hopf algebras in constructing multi-particle Hilbert spaces.\par
~\par
The four spectrum-generating algebras introduced in (\ref{sub}) are spanned by $3$ generators, while
the eight spectrum-generating algebras introduced in (\ref{minimalfer}) and (\ref{minimalpf}) are spanned by $4$ generators.\\
In (\ref{minimalfer}), $f_3^\dagger$ is a primitive element of the graded Lie (super)algebras; similarly, 
$p_3^\dagger$ is a primitive element of the (\ref{minimalpf}) graded Lie (super)algebras. On the other hand, in (\ref{sub}),
$f_3^\dagger = f_1^\dagger f_2^\dagger$ only enters the ${\cal U}(fLS_{sub})$, ${\cal U}(fCLA_{sub})$ Enveloping Algebras (stated otherwise, $f_3^\dagger \notin flS_{sub}$ and $f_3^\dagger \notin fCLA_{sub}$). Similarly, in (\ref{sub}),  $p_3^\dagger=p_1^\dagger p_2^\dagger$ only enters the ${\cal U}(plA_{sub})$, ${\cal U}(pCLS_{sub})$ Enveloping Algebras. \par
~\par
\par
Let us denote as ${\mathfrak g}$ one of the $12$ spectrum-generating algebras defined in (\ref{minimalfer},\ref{minimalpf},\ref{sub}) and as $U\equiv {\cal U}({\mathfrak g})$ its Universal Enveloping Algebra. The two-particle Hilbert space ${\cal H}^{(2)}$ is recovered from the graded Hopf algebra $U$ via application of the $\Delta: U\rightarrow U\otimes U$ coproduct. The graded Hilbert space ${\cal H}^{(2)}$ is a subset of the tensor product of two $4$-dimensional single-particle Hilbert spaces $V$ introduced in (\ref{vspace}):
\bea
{\cal H}^{(2)}&\subset& V\otimes V.
\eea
A hat denotes the action of the coproduct on $V\otimes V$, so that
\bea
&{\widehat{\Delta (U)}}\in End(V\otimes V).&
\eea
We get, in particular,
\bea
{\widehat{\Delta (H_{min})}}&=&H_{min}\otimes {\mathbb I}_4+{\mathbb I}_4\otimes H_{min}\eea
and, for the creation operators,
\bea
{\textrm{either}} \qquad {\widehat{\Delta (f_i^\dagger)}}=f_i^\dagger\otimes {\mathbb I}_4+{\mathbb I}_4\otimes f_i^\dagger&& 
{\textrm{or}} \qquad {\widehat{\Delta (p_i^\dagger)}}=p_i^\dagger\otimes {\mathbb I}_4+{\mathbb I}_4\otimes p_i^\dagger ,
\eea
where $i$ is restricted to be $i=1,2$ from (\ref{sub}), while $i=1,2,3$ for the algebras (\ref{minimalfer}) and (\ref{minimalpf}).\par
The two-particle vacuum state $|vac\rangle^{(2)}$ is the tensor product of two single-particle vacua $|vac\rangle$ introduced in (\ref{fock}):
\bea
|vac\rangle^{(2)} &=& |vac\rangle\otimes |vac\rangle.
\eea
The $V\otimes V$ vector space is spanned by the $16$-component vectors $w_j$ (for $ j =1,2,\ldots, 16$) with entry $1$ in the $j$-th position and $0$ otherwise. Therefore, $|vac\rangle^{(2)}= w_1$. \\ 
The $2$-particle Hilbert space ${\cal H}^{(2)}$ is identified with
\bea
{\cal H}^{(2)} &=& {\widehat{\Delta (U)}}|vac\rangle^{(2)}.
\eea
The $2$-particle Hamiltonian
$H_{min}^{(2)}$ is given by
\bea\label{2phamilt}
H_{min}^{(2)}:={\widehat{\Delta (H_{min})}}.
\eea

{\it Remark 1}: a nilpotent creation operator with grading $\alpha$  satisfies the Pauli exclusion principle provided that  the $\langle \cdot, \cdot\rangle$  bilinear mapping entering the (\ref{gradedbracket}) definition of the graded Lie bracket satisfies the $\langle\alpha,\alpha\rangle =1$ condition. This point is illustrated, by recalling that $(f_1^\dagger)^2=0$, with the $10$-graded $f_1^\dagger$  example. The (\ref{braidedtensor}) braided tensor product implies, for
${\widehat{\Delta((f_1^\dagger)^2)}}$:
\bea 
{\widehat{\Delta((f_1^\dagger)^2)}}&=& {\widehat{\Delta(f_1^\dagger)}}\cdot {\widehat{\Delta(f_1^\dagger)}}=(f_1^\dagger\otimes {\mathbb I}_4+{\mathbb I}_4\otimes f_1^\dagger)\cdot (f_1^\dagger\otimes {\mathbb I}_4+{\mathbb I}_4\otimes f_1^\dagger)=\nonumber\\
&=&f_1^\dagger\otimes f_1^\dagger +({\mathbb I}_4\otimes f_1^\dagger)\cdot(f_1^\dagger\otimes {\mathbb I}_4)=
(1 +(-1)^{\langle 10,10\rangle}) f_1^\dagger\otimes f_1^\dagger.
\eea

The $\langle 10,10\rangle$ value is read from the (\ref{n2cases}) arrays, producing 
\bea
\langle 10,10\rangle =0 &\Rightarrow&{\widehat{ \Delta((f_1^\dagger)^2)}}=2 f_1^\dagger\otimes f_1^\dagger \neq 0 \qquad {\textrm{for $I\equiv LA$ and $III\equiv CLA$,}}\nonumber\\
\langle 10,10\rangle =1 &\Rightarrow& {\widehat{\Delta((f_1^\dagger)^2)}}= 0 \qquad \qquad \qquad\quad  {\textrm{for $II\equiv LS$ and $IV\equiv CLS$.}}
\eea
~\par

{\it Remark 2}: for the fermionic minimal algebras (\ref{minimalfer}) the $2$-particle energy eigenvectors corresponding to the $=\lambda+1$ energy eigenvalue are created by the primitive generator $f_3^\dagger$ satisfying ${\widehat{\Delta (f_3^\dagger)}}=f_3^\dagger\otimes {\mathbb I}_4+{\mathbb I}_4\otimes f_3^\dagger$ and by the composite operator ${\widehat{\Delta (f_1^\dagger\cdot f_2^\dagger)}}$. We get
\bea
\Psi_{E=\lambda+1;\alpha} ~\propto~ {\widehat{\Delta (f_1^\dagger\cdot f_2^\dagger)}}|vac\rangle^{(2)}~~&{\textrm{and}} &~~
\Psi_{E=\lambda+1;\beta} ~\propto~ {\widehat{\Delta (f_3^\dagger)}}|vac\rangle^{(2)}.
\eea
These two eigenvectors are distinct, implying a double degeneracy of the $E=\lambda+1$ energy level.
The same argument is repeated for the parafermionic minimal algebras (\ref{minimalpf}) which present a double degeneracy of the $E=\lambda+1$, $2$-particle, energy level.\par
On the other hand, the absence of the $f_3^\dagger$ ($p_3^\dagger$) generator in $fLS_{sub},~fCLA_{sub}$ (respectively, $pLA_{sub},~ pCLS_{sub}$), implies that the (\ref{sub}) spectrum-generating algebras induce a nondegenerate $E=\lambda+1$, $2$-particle energy level. It follows that the $2$-particle Hilbert spaces defined from (\ref{sub})
have a distinct signature and are physically non-equivalent with respect to the $2$-particle Hilbert spaces recovered from
(\ref{minimalfer},\ref{minimalpf}).\par
~\par
We are now in the position to present the $2$-particle energy spectra and the basis of orthonormal eigenvectors for each one of  the $12$ spectrum-generating algebras introduced in (\ref{minimalfer},{\ref{minimalpf},\ref{sub}).
Once constructed the twelve energy spectra and Hilbert spaces, their mutual relations and physical equivalence/inequivalence are analyzed.\par
For convenience, a list of normalized, $2$-particle, energy eigenvectors obtained from the coproducts of the creation operators is presented. Each one of the $12$ Hilbert spaces induced by  the spectrum-generating algebras is spanned by a subset of the eigenvectors presented in the following list. 
At a given energy level $E$, corresponding normalized eigenvectors are given by:
\bea
&\begin{array}{lll}
E=0: &\quad \Psi_0 &=~ w_1,\\
E=1: &\quad  \Psi_1 &=~ \frac{1}{\sqrt 2}(w_3+w_9),\\
E=\lambda: &\quad \Psi_\lambda &=~ \frac{1}{\sqrt 2}(w_2+w_5),\\
E={\lambda+1}: &\quad  \Psi_{\lambda+1; \alpha_\pm}& =~ \frac{1}{\sqrt 2}(w_7\pm w_{10}),\\
E=\lambda+1: &\quad \Psi_{\lambda+1;\beta} &=~ \frac{1}{\sqrt 2}(w_4+w_{13}),\\
E=2: &\quad  \Psi_2 &=~ w_{11}\\
E=2\lambda: &\quad \Psi_{2\lambda} &=~ w_6\\
E=\lambda+2: &\quad  \Psi_{\lambda+2;\pm} &=~ \frac{1}{\sqrt 2}(w_{12}\pm w_{15}),\\
E=2\lambda+1: &\quad \Psi_{2\lambda+1; \pm} &=~ \frac{1}{\sqrt 2}(w_8\pm w_{14}),\\
E=2\lambda+2: &\quad  \Psi_{2\lambda+2} &=~ w_{16}.
\end{array}&
\eea

\subsection{$2$-particle, $3$-generator, spectrum-generating algebras}

The energy spectra and the $2$-particle Hilbert spaces
${\cal H}^{(2)}$ induced by the (\ref{sub}) spectrum-generating algebras are:
\par~\\
{\it i}) ~ For $fLS_{sub}$, the Hilbert space  is spanned by $4$ eigenvectors;
\par the energy spectrum is given by $E=0,1,\lambda,\lambda+1$;\par
the corresponding energy eigenvectors are $\Psi_0, ~\Psi_1,~\Psi_\lambda$ and
\bea\label{lp1}
\frac{1}{\sqrt 2}( \Psi_{\lambda+1;\beta}- \Psi_{\lambda+1; \alpha_-}) &=&\frac{1}{\sqrt 4}(w_4-w_7+w_{10}+w_{13}) .
\eea
~\\
{\it ii}) ~For $pCLS_{sub}$, the Hilbert space  is spanned by $4$ eigenvectors;
\par the energy spectrum is given by $E=0,1,\lambda,\lambda+1$;\par
the corresponding energy eigenvectors are $\Psi_0, ~\Psi_1,~\Psi_\lambda$ and
\bea\label{lp2}
\frac{1}{\sqrt 2}( \Psi_{\lambda+1;\beta}+\Psi_{\lambda+1; \alpha_+}) &=&\frac{1}{\sqrt 4}(w_4+w_7+w_{10}+w_{13}) .
\eea
Due to the $\pm w_7$ sign entering (\ref{lp1},\ref{lp2}) the Hilbert spaces induced by $fLS_{sub}$, $pCLS_{sub}$ differ; nevertheless, as discussed in Subsection {\bf 5.3}, they are physically equivalent.\par
~\\

{\it iii}) ~ For $fCLA_{sub}$, the Hilbert space  is spanned by $9$ eigenvectors;
\par the energy spectrum is given by $E=0,1,\lambda,\lambda+1, 2, 2\lambda, \lambda+2, 2\lambda+1, 2\lambda+2$;\par
the energy eigenvectors are split into the six eigenvectors
\bea\label{common1}
&\Psi_0, ~\Psi_1,~\Psi_\lambda,~\Psi_2,~\Psi_{2\lambda},~ \Psi_{2\lambda+2}&
\eea 

and the three extra, sign-dependent, eigenvectors
\bea\label{extra1}
\frac{1}{\sqrt 2}( \Psi_{\lambda+1;\beta}- \Psi_{\lambda+1; \alpha_-}) &=&~\frac{1}{\sqrt 4}(w_4-w_7+w_{10}+w_{13}) ,\nonumber\\
\Psi_{\lambda+2;-}&=&~ \frac{1}{\sqrt 2}(w_{12}- w_{15}),\nonumber\\
\Psi_{2\lambda+1;-}&=&~ \frac{1}{\sqrt 2}(w_{8}- w_{14}).
\eea

~\\
{\it iv}) ~For $pLA_{sub}$, the Hilbert space  is spanned by $9$ eigenvectors;
\par the energy spectrum is given by $E=0,1,\lambda,\lambda+1, 2, 2\lambda, \lambda+2, 2\lambda+1, 2\lambda+2$;\par
the energy eigenvectors are split into the six eigenvectors
\bea\label{common2}
&\Psi_0, ~\Psi_1,~\Psi_\lambda,~\Psi_2,~\Psi_{2\lambda},~ \Psi_{2\lambda+2}&
\eea

and the three extra, sign-dependent, eigenvectors
\bea\label{extra2}
\frac{1}{\sqrt 2}( \Psi_{\lambda+1;\beta}+ \Psi_{\lambda+1; \alpha_+}) &=&~\frac{1}{\sqrt 4}(w_4+w_7+w_{10}+w_{13}) ,\nonumber\\
\Psi_{\lambda+2;+}&=&~ \frac{1}{\sqrt 2}(w_{12}+ w_{15}),\nonumber\\
\Psi_{2\lambda+1;+}&=&~ \frac{1}{\sqrt 2}(w_{8}+ w_{14}).
\eea

As discussed in Subsection {\bf 5.3}, the different signs entering (\ref{extra1},\ref{extra2}) allow to discriminate the Hilbert space induced by $fCLA_{sub}$ from the Hilbert space induced by $pLA_{sub}$. A signature of the colored parastatistics is encoded in the signs entering $\Psi_{\lambda+2;\pm}$ and $\Psi_{2\lambda+1;\pm}$.

\subsection{$2$-particle, $4$-generator, spectrum-generating algebras}

For the $4+4$ minimal Lie (super)algebras, the  corresponding $2$-particle Hilbert spaces ${\cal H}^{(2)}$ obtained from the fermionic (\ref{minimalfer}) and the parafermionic (\ref{minimalpf}) constructions are identical. We can therefore set
\bea
{\cal H}^{(2)}_{LS}:= {\cal H}^{(2)}_{fLS}= {\cal H}^{(2)}_{pLS}, &\quad& {\cal H}^{(2)}_{CLS}:= {\cal H}^{(2)}_{fCLS}= {\cal H}^{(2)}_{pCLS},\nonumber\\  {\cal H}^{(2)}_{LA}:= {\cal H}^{(2)}_{fLA}= {\cal H}^{(2)}_{pLA}, &\quad&{\cal H}^{(2)}_{CLA}:= {\cal H}^{(2)}_{fCLA}= {\cal H}^{(2)}_{pCLA}.
\eea

Their energy spectra and spanning eigenvectors are given by:\par~\\
{\it v}) ~For $LS_{min}$, the Hilbert space ${\cal H}^{(2)}_{LS}$ is spanned by $8$ eigenvectors;
\par the energy spectrum is given by $E=0,1,\lambda,\lambda+1, \lambda+2, 2\lambda+1, 2\lambda+2$, with the $\lambda+1$ eigenvalue being double degenerate;\par
the energy eigenvectors are split into the five eigenvectors
\bea&\Psi_0, ~\Psi_1,~\Psi_\lambda, ~\Psi_{\lambda+1;\beta}, ~\Psi_{2\lambda+2}&
\eea
and the three extra, sign-dependent, eigenvectors
\bea\label{lp5}
&
 \Psi_{\lambda+1;\alpha_-}=\frac{1}{\sqrt 2}(w_{7}-w_{10}), \quad
 \Psi_{\lambda+2;+}=\frac{1}{\sqrt 2}(w_{12}+w_{15}),\quad 
 \Psi_{2\lambda+1;+}=\frac{1}{\sqrt 2}(w_8+w_{14}); &
\eea
the vectors $\Psi_{\lambda+1;\alpha_-}, ~\Psi_{\lambda+1;\beta}$ span the two-dimensional, degenerate, 
$(\lambda+1)$-eigenspace.\par
~\\
{\it vi}) ~For $CLS_{min}$, the Hilbert space ${\cal H}^{(2)}_{CLS}$ is also spanned by $8$ eigenvectors, with the same energy spectrum given by
\par $E=0,1,\lambda,\lambda+1, \lambda+2, 2\lambda+1, 2\lambda+2$ and the $\lambda+1$ eigenvalue being double degenerate;\par
the energy eigenvectors are split into the five eigenvectors
\bea&\Psi_0, ~\Psi_1,~\Psi_\lambda, ~\Psi_{\lambda+1;\beta}, ~\Psi_{2\lambda+2}&
\eea
and the three extra, sign-dependent, eigenvectors
\bea\label{lp6}
&
 \Psi_{\lambda+1;\alpha_+}=\frac{1}{\sqrt 2}(w_{7}+w_{10}), \quad
 \Psi_{\lambda+2;-}=\frac{1}{\sqrt 2}(w_{12}-w_{15}),\quad 
 \Psi_{2\lambda+1;-}=\frac{1}{\sqrt 2}(w_8-w_{14}); &
\eea
the vectors $\Psi_{\lambda+1;\alpha_+}, ~\Psi_{\lambda+1;\beta}$ span the two-dimensional, degenerate, 
$(\lambda+1)$-eigenspace.\par
~\\
As discussed in Subsection {\bf 5.3}, the different signs entering (\ref{lp5},\ref{lp6}) allow to discriminate the Hilbert space induced by $LS_{min}$ from the Hilbert space induced by $CLS_{min}$.\par
~\\
{\it vii}) ~For $LA_{min}$, the Hilbert space ${\cal H}^{(2)}_{LA}$ is spanned by $10$ eigenvectors;
\par the energy spectrum is given by $E=0,1,\lambda,\lambda+1, 2, 2\lambda, \lambda+2, 2\lambda+1, 2\lambda+2$, with the $\lambda+1$ eigenvalue being double degenerate;\par
the energy eigenvectors are split into the seven eigenvectors
\bea&\Psi_0, ~\Psi_1,~\Psi_\lambda, ~\Psi_{\lambda+1;\beta},~\Psi_2,~\Psi_{2\lambda}, ~\Psi_{2\lambda+2}&
\eea
and the three extra, sign-dependent, eigenvectors
\bea\label{lp7}
&
 \Psi_{\lambda+1;\alpha_+}=\frac{1}{\sqrt 2}(w_{7}+w_{10}), \quad
 \Psi_{\lambda+2;+}=\frac{1}{\sqrt 2}(w_{12}+w_{15}),\quad 
 \Psi_{2\lambda+1;+}=\frac{1}{\sqrt 2}(w_8+w_{14}); &
\eea
the vectors $\Psi_{\lambda+1;\alpha_+}, ~\Psi_{\lambda+1;\beta}$ span the two-dimensional, degenerate, 
$(\lambda+1)$-eigenspace.\par
~\\
{\it viii}) ~For $CLA_{min}$, the Hilbert space ${\cal H}^{(2)}_{CLA}$ is also spanned by $10$ eigenvectors;
\par the energy spectrum is given by $E=0,1,\lambda,\lambda+1, 2, 2\lambda, \lambda+2, 2\lambda+1, 2\lambda+2$, with the $\lambda+1$ eigenvalue being double degenerate;\par
the energy eigenvectors are split into the seven eigenvectors
\bea&\Psi_0, ~\Psi_1,~\Psi_\lambda, ~\Psi_{\lambda+1;\beta},~\Psi_2,~\Psi_{2\lambda}, ~\Psi_{2\lambda+2}&
\eea
and the three extra, sign-dependent, eigenvectors
\bea\label{lp8}
&
 \Psi_{\lambda+1;\alpha_-}=\frac{1}{\sqrt 2}(w_{7}-w_{10}), \quad
 \Psi_{\lambda+2;-}=\frac{1}{\sqrt 2}(w_{12}-w_{15}),\quad 
 \Psi_{2\lambda+1;-}=\frac{1}{\sqrt 2}(w_8-w_{14});&
\eea
the vectors $\Psi_{\lambda+1;\alpha_-}, ~\Psi_{\lambda+1;\beta}$ span the two-dimensional, degenerate, 
$(\lambda+1)$-eigenspace.\par
~\\
As discussed in Subsection {\bf 5.3}, the different signs entering (\ref{lp7},\ref{lp8}) allow to discriminate the Hilbert space induced by $LA_{min}$ from the Hilbert space induced by $CLA_{min}$.
\subsection{Discriminating Hilbert spaces via yes/no projection measurements}

In the $2$-particle sectors the $12$ spectrum-generating algebras (\ref{minimalfer},\ref{minimalpf},\ref{sub}) induce $8$ different Hilbert spaces (itemized from {\it i} to {\it viii} in the Subsections {\bf 5.1},{\bf 5.2}). These Hilbert spaces are split into $4$ pairs which produce distinct signatures of inequivalent physics; the signature is directly read from the energy spectrum of the associated $2$-particle quantum models. The dimensions  of the $2$-particle Hilbert spaces produced by each pair are:
\bea\label{pairs}
{\textrm{$4$-dim. ~Hilbert space~:}}  &~&  fLS_{sub} ~~ \leftrightarrow~~ pCLS_{sub},\nonumber\\
{\textrm{$8$-dim. ~Hilbert space~:}}  &~& ~LS_{min} ~~ \leftrightarrow~~ ~CLS_{min},\nonumber\\
{\textrm{$9$-dim. ~Hilbert space~:}}  &~&  fLA_{sub} ~~ \leftrightarrow~~ pCLA_{sub},\nonumber\\
{\textrm{$10$-dim. ~Hilbert space~:}}  &~&  ~LA_{min} ~~ \leftrightarrow~~ ~CLA_{min}.
\eea

The four pairs are easily discriminated:\par
~\\
{\it a}) An $E=2$ energy eigenstate is present, e.g., in the $fLA_{sub}\leftrightarrow pCLA_{sub}$ and 
$LA_{min} \leftrightarrow CLA_{min}$ pairs; on the other hand this eigenstate is absent, due to the Pauli exclusion principle, from the
superalgebra pairs $fLS_{sub} \leftrightarrow pCLS_{sub}$ and $LS_{min} \leftrightarrow CLS_{min}$.\par
~\\
{\it b}) The pairs $fLA_{sub}\leftrightarrow pCLA_{sub}$ versus $LA_{min} \leftrightarrow CLA_{min}$ are discriminated, due to the degeneracy of the $(\lambda+1)$ energy eigenspace (respectively $1$ for the first pair and $2$ for the second pair).\par
~\\
{\it c}) The pairs $fLS_{sub} \leftrightarrow pCLS_{sub}$ versus $LS_{min} \leftrightarrow CLS_{min}$ are discriminated for the presence, in the latter case, of extra energy eigenstates (for instance, at $E=\lambda+2$).\par
~\par 
It follows that the four pairs are all physically inequivalent; their inequivalence is not related to parastatistics. In order to detect a signature of parastatistics one has to determine, {\it within each pair}, a possible physical inequivalence.
Indeed, each pair couples a Hilbert space obtained from  an ordinary Lie (super)algebra (that is, ordinary bosons/fermions), with a Hilbert space obtained from a color Lie (super)algebra (i.e., implying colored paraparticles).\par
Since the differences in the eigenvectors basis are at most determined by a sign, it makes sense to search for yes/no projection measurements which spot the difference of an ordinary Lie (super)algebra eigenstate versus its corresponding colored partner. Let's now proceed.
\par
~\\
{\it First scenario}: $fLS_{sub}$ versus $pCLS_{sub}$  (i.e., ${\cal H}^{(2)}_{fLS_{sub}}$ versus ${\cal H}^{(2)}_{pCLS_{sub}}$ Hilbert spaces). \\
The only difference is in the normalized $(\lambda+1)$-eigenvector, given by
\bea
\frac{1}{\sqrt 4}(w_4\pm w_7+w_{10}+w_{13}), &&{\textrm{where the sign is $+$ for $fLS_{sub}$ and $-$ for $pCLS_{sub}$.}}
\eea
In the $4$-dimensional subspace 
\bea 
 w_\pm &:=&\frac{1}{\sqrt 4}(w_4, w_{10}, w_{13}, \pm w_7)^T\propto (1,1,1,\pm 1)^T
\eea 
one should look for $P_\pm^{4d}$ hermitian projectors satisfying the conditions
\bea\label{4dconditions}
&&(P_{\pm}^{4d})^2 = (P_{\pm}^{4d}), \qquad (P_{\pm}^{4d})(P_{\mp}^{4d})=(P_{\mp}^{4d})(P_{\pm}^{4d})=0, \qquad (P_{\pm}^{4d}+P_{\mp}^{4d})= {\mathbb I}_4, \nonumber\\
&&(P_{\pm}^{4d})^\dagger = (P_{\pm}^{4d}), \qquad (P_{\pm}^{4d}) w_\pm =\pm w_\pm, \qquad (P_{\pm}^{4d}) w_\mp = 0.
\eea
It is easily checked, by trying to impose the (\ref{4dconditions}) conditions on the most general $4\times 4$ matrices $P_{\pm}^{4d}$, that the whole set of (\ref{4dconditions}) conditions is incompatible, leading to a contradiction. No such $P_{\pm}^{4d}$ hermitian projectors, discriminating $w_+$ from $w_-$, can be found. It follows that, even if as Hilbert spaces ${\cal H}^{(2)}_{fLS_{sub}}$ and ${\cal H}^{(2)}_{pCLS_{sub}}$ are different, they are physically equivalent. Symbolically:
\bea\label{firstpair}
{\cal H}^{(2)}_{fLS_{sub}}\neq {\cal H}^{(2)}_{pCLS_{sub}},  &\qquad & {\cal H}^{(2)}_{fLS_{sub}}\equiv {\cal H}^{(2)}_{pCLS_{sub}}. 
\eea
\par
~\\
{\it Discriminating the three remaining pairs}: the situation is different for the three remaining pairs, namely $LS_{min}$ versus  $CLS_{min}$, next  $fLA_{sub}$ versus $pCLA_{sub}$ and, finally,  $LA_{min}$ versus  $CLA_{min}$. Unlike the previous (\ref{firstpair}) result,  for these three pairs a yes/no projection measurement can discriminate the ordinary case from the colored case. Since the mechanism at work for the three pairs is the same, it will be illustrated just for the  $LS_{min}\leftrightarrow CLS_{min}$ comparison.
\par
~\par
For the induced $LS_{min}$ versus $CLS_{min}$ Hilbert spaces, at the energy level $E=\lambda +2$ (the  $E=2\lambda+1$ eigenvalue is similarly treated) the corresponding eigenvectors are
 \bea 
\Psi_{\lambda+2;\pm}=\frac{1}{\sqrt 2}(w_{12}\pm w_{15}),&&{\textrm{where the sign is $+$ for $LS_{min}$ and $-$ for $CLS_{min}$.}}
\eea
One should now search, in the $2$-dimensional subspace 
\bea 
 w_{\lambda+2;\pm} &:=&\frac{1}{\sqrt 4}(w_{12},  \pm w_{15})^T\propto (1,\pm 1)^T,
\eea 
for $P_\pm^{2d}$ hermitian projectors satisfying the conditions
\bea\label{2dconditions}
&&(P_{\pm}^{2d})^2 = (P_{\pm}^{2d}), \qquad (P_{\pm}^{2d})(P_{\mp}^{2d})=(P_{\mp}^{2d})(P_{\pm}^{2d})=0, \qquad (P_{\pm}^{2d}+P_{\mp}^{2d})= {\mathbb I}_2, \nonumber\\
&&(P_{\pm}^{2d})^\dagger = (P_{\pm}^{2d}), \qquad (P_{\pm}^{2d}) w_{\lambda+2;\pm}=\pm w_{\lambda+2;\pm}, \qquad (P_{\pm}^{2d}) w_{\lambda+2;\mp} = 0.
\eea
An obvious solution to the above conditions are the $2\times 2$ projectors
{\footnotesize{\bea\label{2dproj}
P_{\pm}^{2d}&=&\frac{1}{2}\left(\begin{array}{cc} 1&\pm 1\\\pm 1&1\end{array}\right).
\eea}}
In the $LS_{min} \leftrightarrow CLS_{min}$ comparison, once selected the system in the $E=\lambda+2$ energy eigenvalue, the output of a projection measurement can determine whether the state under consideration belongs to an ordinary Hilbert space or to a colored Hilbert space.
The same analysis can be repeated for the $E=\lambda+2$ energy eigenvalues of $fLA_{sub}$ versus $pCLA_{sub}$ and of $LA_{min}$ versus $CLA_{min}$. Symbolically, the physical inequivalence of these Hilbert spaces can be expressed as
\bea
&{\cal H}^{(2)}_{LS_{min}}\not\equiv {\cal H}^{(2)}_{CLS_{min}},  \qquad  {\cal H}^{(2)}_{fLA_{sub}}\not\equiv {\cal H}^{(2)}_{pCLA_{sub}},\qquad {\cal H}^{(2)}_{LA_{min}}\not\equiv {\cal H}^{(2)}_{CLA_{min}}.&
\eea
Projector operators like $P_{\pm}^{2d} $ can naturally be embedded in $16\times 16$ observables of the $2$-particle Hilbert space. The observables are hermitian operators which further satisfy the condition of being $00$-graded. This requirement is due to the results of the measurements being real numbers and not more general graded functions. The nonvanishing entries of a $16\times 16$ observable can only be accommodated in the matrix given in (\ref{2particleobs}).

\section{Outcomes of the minimal Gedankenexperiment}

The results derived in the previous Section are here summarized; their implications, for a minimal Gedankenexperiment  to detect paraparticles which are exchanged via the permutation group, are analyzed.\par
The pairs of fermionic oscillators and ${\mathbb Z}_2^2$-graded parafermionic oscillators, respectively introduced in (\ref{hfer}) and (\ref{hpf}), induce $12$ spectrum-generating algebras.  They correspond to all possible Lie (super)algebras, either colored or not, which are compatible with the matrix realizations of the fermionic and parafermionic oscillators. The spectrum-generating algebras are recovered from the  four admissible assignments of $2$-bit (para)statistics presented in (\ref{n2cases}). Following the terminology of \cite{nbits}, the  different assignments implement an ``algebraic statistical transmutation" of the given creation/annihilation operators. \par
The $12$ spectrum-generating algebras are presented in (\ref{minimalfer},{\ref{minimalpf},\ref{sub}). In the $2$-particle sector of the minimal quantum model (\ref{2phamilt}), $8$ out of the $12$ spectrum-generating algebras
($fLS_{sub}$, $pCLS_{sub}$, $LS_{min}$, $CLS_{min}$, $fLA_{sub}$, $pCLA_{sub}$, $LA_{min}$, $CLA_{min}$)
define different Hilbert spaces, which are paired according to (\ref{pairs}). 
For each such pair, an ordinary Lie (super)algebra is paired with a Rittenberg-Wyler ${\mathbb Z}_2^2$-graded color Lie (super)algebra inducing the same energy spectrum.\\
Apart from the $fLS_{sub}\leftrightarrow pCLS_{sub}$ pair, in the remaining three cases ($LS_{min}\leftrightarrow CLS_{min}$, $fLA_{sub}\leftrightarrow pCLA_{sub}$, $LA_{min}\leftrightarrow CLA_{min}$) a yes/no projection measurement on a given eigenstate can discriminate  ordinary versus color Lie (super)algebra in the given pair.
\par
~\par 
Let's discuss the situation in more detail. Let's say that we have prepared a $2$-particle quantum system and determined its energy spectrum. The absence, e.g., of an $E=2$ energy eigenstate, implied by the  $fLA_{sub}, ~ pCLA_{sub},~LA_{min},~CLA_{min}$ algebras, excludes their induced Hilbert spaces\footnote{The same analysis can be repeated and reversed if, for instance, the $E=2$ energy eigenstate is present.}. Furthermore, the presence, e.g., of a $E=\lambda+2$ energy eigenstate absent in $fLS_{sub}, ~pCLS_{sub}$ only leaves the possibility of the Hilbert spaces
induced by $LS_{min}\leftrightarrow CLS_{min}$. \par

The $2$-component $(\lambda+2)$-eigenstate $\Psi_{\lambda+2;\pm}= \frac{1}{\sqrt 2}(w_{12}\pm w_{15})$ differs by a sign (it is $+1$ for $LS_{min}$ and $-1$ for $CLS_{min}$). This difference can be spotted by applying a yes/no projection measurement defined by a $P_+^{2d}$ projector, as presented in (\ref{2dproj}). 
The implication is that the Rittenberg-Wyler  \cite{{rw1},{rw2}} ${\mathbb Z}_2^2$-graded color Lie (super)algebras present physical consequences which cannot be reproduced by ordinary bosons/fermions.

\subsection{A Gedankenexperiment scenario}

For a better understanding of the physical consequences, the following scenario can be considered. Two experimentalists conduct a blind test in separate rooms. The first experimentalist produces $2$-particle energy eigenstates (possibly controlling some input parameters or turning on/off some switches); the second experimentalist measures the output. They have determined that the energy spectrum corresponds to the $LS_{min}$/${CLS}_{min}$ Hilbert spaces. A which is which's refinement (the $LS_{min}$ versus the $CLS_{min}$ Hilbert space) requires the yes/no measurement
associated with the projector operator; the projection is applied to the $E=\lambda+2$ energy eigenstate.\par
At this stage an ambiguity can be noted. The $P_+^{2d}, P_-^{2d}$ projectors from (\ref{2dproj}) give complementary results, making unclear the proper interpretation of the projection measurement. Is it a $+1$ eigenvalue associated with ordinary particles or with colored paraparticles?\par
~\par In \cite{ijgmmp} I discussed in detail the solution of this ambiguity. The measurement device associated with the projector needs to be calibrated. In a series of measurements, the first measurement acts as a calibration test.
Once calibrated the measuring apparatus, from the second measurement on, the actual results of the test are collected.\par
Since the aim is to find an unambiguous signal of parastatistics, the {\it conventionality's argument} can be applied to the calibration test. It postulates that the ordinary bosons/fermions statistics is the result of the calibration measurement. Once accepted this, an inequivocal signature of parastatistics is found if some (theoretically, at least one) of the following measurements produce opposite outputs with respect to the calibration test.\par
~\par
This is the minimal setup to detect the presence of paraparticles which, in this case, obey a Rittenberg-Wyler's colored parastatistics. More ``spectacular" theoretical signatures of parastatistics have also been obtained; for instance, in cases investigated in \cite{nbits}, a signature of parastatistics is directly read from the energy spectrum of multi-particle models. The minimal scenario here presented is attractive for the potentialities which it offers; its simple logical rules (once properly encoded) can be used as guidelines for realizing an actual experimental test of detecting paraparticles in a laboratory. 

\subsection{Towards an experimental chirality test of paraparticles}

To understand what should be done to realize an actual experiment, we can compare the proposed chirality test with the famous
experimental discovery by Madame Wu \cite{wuparity} of parity violation in weak interactions. That experiment corroborated  the Lee-Yang's proposed theoretical mechanism \cite{leya} to solve the  $\tau-\theta$ puzzle  in the kaon's decay. One can note a similarity and a difference with respect to the proposed minimal scenario for detecting paraparticles exchanged via the permutation group.\par
{{\it About the similarity}}: both cases deal with a {\it chirality test}; in the minimal Gedankenexperiment the problem of detecting paraparticles is mapped, see (\ref{2dproj}), in determining the $\varepsilon = \pm 1$ parity
of a projection measurement 

{\footnotesize{\bea\label{proj}
\frac{1}{2}\left(\begin{array}{cc} 1& 1\\ 1&1\end{array}\right)\left(\begin{array}{c} 1\\ \varepsilon\end{array} \right)&=&\frac{1}{2}(1+\varepsilon)\left(\begin{array}{c} 1\\ \varepsilon\end{array} \right).
\eea}}

{\it About the difference}: in the case of weak interactions, a well-defined Lagrangian was put to test; in the case of paraparticles, on the other hand, the minimal scenario presents a well-defined logical scheme. This logical scheme proves that Rittenberg-Wyler's colored paraparticles are theoretically detectable. In order to perform an actual experimental test, the logical setup should be properly encoded and ``translated" into a doable physical experiment. This is an open challenge. \par
~\par
In order to make contact with the possibility of engineering paraparticles, in \cite{nbits} the  (\ref{n2cases}) arrays of graded color Lie (super)algebras have been presented in terms of Boolean logic gates. This is a first step towards connecting abstract mathematical structures to an operative physical framework. The following points can be itemized:\par
- The solution of the {\it calibration's ambiguity} requires a mechanism which allows to produce or engineer both ordinary particles and colored paraparticles; in this way the needed opposite outputs from the whole series of projection measurements are obtained.\par
- 
The comparison of
\bea\label{comparison12}
LS_{min}~~  {\textrm{versus}} ~~  CLS_{min} &\qquad{\textrm{and/or}}\qquad & 
LA_{min}~~  {\textrm{versus}} ~~CLA_{min} 
\eea
requires the creation operator $f_3^\dagger$ (respectively, $p_3^\dagger$) to be encoded as a {\it primitive element} of the spectrum-generating algebra. \par
- On the other hand, the comparison of
\bea\label{comparison3}
 &fLA_{sub}\quad{\textrm{versus}}\quad pCLA_{sub}&
\eea
requires $f_3^\dagger$ (respectively, $p_3^\dagger$) to be implemented only as an element of the corresponding Enveloping Algebra generated by $f_1^\dagger, f_2^\dagger$ (respectively, $p_1^\dagger, p_2^\dagger$).\par
No matter which of the three comparative scenarios  from (\ref{comparison12},\ref{comparison3}) is picked up, the colored paraparticles are discriminated 
from ordinary paraparticles via a (\ref{proj}) projection measurement applied on a suitable energy eigenstate.\par
~\par
On the experimental side some techniques to engineer paraparticles in the laboratory have been developed. In particular, in \cite{{parasim},{paraexp}} this possibility has been proved by using the spin of a trapped atomic ion and two of its bosonic modes of motion in the trap, with tailoring laser-induced couplings between them.  This result does not yet guarantee that a signature of paraparticles has been experimentally detected. What is still lacking it is the proof that the engineered, laser-induced models, could not be reproduced by ordinary bosons/fermions statistics. This is why it is important that an experiment is conceived to test a distinct signature of parastatistics (as the ones presented in \cite{{top1},{top2},{waha},{nbits}}).\par
A different technique to detect parastatistics is based on photonic. In \cite{photonic} the authors show how to observe fermionic statistics (and also intermediate anyonic statistics) via entanglement of photons.\par
Perhaps the most promising technique to engineer Rittenberg-Wyler's colored paraparticles makes use of {\it qudits}, the generalization of ordinary qubits described by a $d$-dimensional Hilbert space.
It is detailed, in the  \cite{qudits} review article, that qudits provide, for $d>2$, a larger space than qubits to store and process information. Physical realizations for qudit computation include, besides  the already mentioned photonic platforms and ion traps, nuclear magnetic resonance. Universal Qudit Gates realize qudit-based quantum computing and information processing. When specialized to $d=4$, the qudits are also known as {\it ququarts}. At $d=4$, the $d\times d$ matrices acting on qudits are given by $4\times 4$ matrices, as the ones entering (\ref{gradedmatrices}) and possessing a ${\mathbb Z}_2^2$ grading. It is therefore quite tempting to use the already available technology to manipulate qudits for constructing and detecting ${\mathbb Z}_2^2$-graded colored paraparticles. An investigation along these lines could lead to the first experimental signature of
permutation-group parastatistics.

\section{Conclusions and perspectives}

The paper presents a minimal Gedankenexperiment to detect a signature of paraparticles exchanged via the permutation group.
The main aim is to fill a gap between the known theoretical signatures presented in certain \cite{{top1},{top2},{nbits},{waha}} quantum models and the (still open) challenge to realize an actual experiment to discriminate paraparticles from ordinary bosons/fermions.\par
~\\
The following question is answered: \par
{\it Which is the minimal theoretical setting to detect paraparticles?}\\
The minimal scenario to introduce a parastatistics is given by the Rittenberg-Wyler's colored paraparticles \cite{{rw1},{rw2}}.
The paper revises and simplifies the \cite{{top1},{top2}}
tests, proving that the distinction between ordinary particles and ${\mathbb Z}_2^2$-graded colored paraparticles can be mapped into a {\it parity} (also referred as {\it chirality}) measurement: a yes/no projection measure applied to a given energy eigenstate. Subtle issues, like the solution of the {\it calibration's ambiguity} in a series of parity measurements, are discussed in the Gedankenexperiment scenario of Section {\bf 6}. \par
~\par
The overall picture is that of an unambiguous set of logical rules; for a doable experiment, these logical rules have to be concretely implemented. It is argued that the most promising candidate for the experiment seems to be provided by the quantum technology based on manipulating the $d$-dimensional qudits via Universal Qudits Gates; the minimal vector space of ${\mathbb Z}_2^2$-graded paraparticles is indeed $d=4$ dimensional.\par

~\par
In recent years the fields of parastatistics experienced a boost: more refined theoretical considerations, new experimental possibilities from condensed matter and quantum information technology, with technological applications starting to be seriously considered.  \par
Advances in recent months include, e.g., refinements of the notions of {\it permutation invariance}; if satisfied, both requirements of ``complete invariance"  and ``quantum permutation invariance" introduced in \cite{iqoqi} rule out parastatistics based on higher dimensional representations of the permutation group (for an account of this parastatistics, see \cite{gold}).  It is interesting to note that the Rittenberg-Wyler's $n$-bit colored paraparticles of \cite{{top1},{top2},{nbits}} not only evade the locality principles of \cite{conventionality}, but also the stringent notions introduced in \cite{iqoqi}. Indeed, the $n$-bit colored paraparticles are accommodated in  ({\it differenty organized} with respect to ordinary bosons/fermions) unidimensional reps of the permutation group.\par
An interesting technological application of ${\mathbb Z}_2^2$-graded paraparticles has been very recently proposed in \cite{qcz2z2}. It allows implementing a procedure for error correction in photonic platforms. 
The paper further discusses possible future paraparticle-based quantum computers using graded qudits and ququarts. As already mentioned, this is the framework which could turn the minimal Gedankenexperiment into an actual experimental detection of ${\mathbb Z}_2^2$-graded paraparticles.\par
~
\\ {\Large{\bf Acknowledgments}}
{}~\par{}~\\
I am indebted to Zhanna Kuznetsova for many years of discussions and elucidations on parastatistics and the role of colored Lie (super)algebras. Rodrigo G. Rana, who proposed implementing colored Lie (super)algebras via Boolean logic gates, deserves special thanks. I have profited of several pleasant discussions and enlightening exchanges of ideas with Gerald G. Goldin. At IQOQI in Vienna I had very enjoyable and illuminating discussions on the different notions of permutation invariance with Thomas D. Galley, Manuel Mekonnen and Markus P. M\"uller. I am grateful to Hubert de Guise who introduced me to the notion of qudits, ququarts and the possibilities they offer. I am grateful to Pengming Zhang for the hospitality at Sun Yat-sen University in Zhuhai where this work was concluded.\par
This work was supported by CNPq (PQ grant 308846/2021-4).

\end{document}